\documentclass[11pt,a4paper]{article}
\usepackage{amsfonts}
\usepackage{amsmath}
\usepackage{amssymb}
\usepackage[utf8]{inputenc}
\usepackage[pdftex]{color,graphicx,hyperref}
\usepackage{caption}
\usepackage{subcaption}
\usepackage{soul}
\usepackage{url}
\usepackage{setspace}
\usepackage{nameref}

\begin{document}
\title{ \bf Comparing the hierarchy of keywords in on-line news portals}
\author{Gergely Tib{\'e}ly$^{1,*}$, David Sousa-Rodrigues$^{2}$, P{\'e}ter Pollner$^3$ \\ and Gergely Palla$^3$\\
\footnotesize{$^1$Dept. of Biological Physics, E{\"o}tv{\"os} University, H-1117 Budapest, Hungary}\\
\footnotesize{$^*$tibelyg@hal.elte.hu}\\
\footnotesize{$^2$The Design Group, Faculty of Maths, Computing and Technology}\\
\footnotesize{ The Open University, Walton Hall, Milton Keynes, MK7 6AA United Kingdom }\\
\footnotesize{$^3$MTA-ELTE Statistical and Biological Physics Research Group,}\\
\footnotesize{Hungarian Academy of Sciences, H-1117 Budapest, Hungary,}} 
\maketitle

\begin{abstract}
The tagging of on-line content with informative keywords is a wide\-spread phenomenon from scientific article repositories through blogs to on-line news portals. In most of the cases, the tags on a given item are free words chosen by the authors independently. Therefore, relations among keywords in a collection of news items is unknown. However, in most cases the topics and concepts described by these keywords are forming a latent hierarchy, with the more general topics and categories at the top, and more specialised ones at the bottom. Here we apply a recent, cooccurrence-based tag hierarchy extraction method to sets of keywords obtained from four different on-line news portals. The resulting hierarchies show substantial differences not just in the topics rendered as important (being at the top of the hierarchy) or of less interest (categorised low in the hierarchy), but also in the underlying network structure. This reveals discrepancies between the plausible keyword association frameworks in the studied news portals. 
\end{abstract}

\newpage

\section{Introduction}

Hierarchical organisation is a widespread phenomenon in nature and society. Signs of hierarchy were recorded in various animal flocks \cite{Huber_crayfish,Tamas_pigeons,Pigeon_context,McCowan_macaque}, in social interactions \cite{Guimera_hier_soc,our_pref_coms,Sole_hier_soc}, in urban planning \cite{Krugman_urban,Batty_urban}, in ecological systems \cite{Hirata_eco,Wickens_eco} and in evolution \cite{Eldrege_book,McShea_organism}. Since a natural representation of hierarchies is given by directed acyclic graphs, hierarchical organisation became a very relevant concept also in complex network theory \cite{Laci_hier_scale,Sneppen_hier_measures,Newman_hier,Pumain_book,Sole_chaos_hier,Enys_hierarchy,Sole_hier_PNAS,Kaiser_neural,Zeng_Ecoli}.

The association of tags to various on-line contents have became ubiquitous, as various tags may indicate the topic of news-portal feeds and blog post, the genre of films or music records on file sharing portals, or the kind of goods offered in Web stores. These tags usually serve as keywords, providing a rough description of the given entity, helping the users in a fast decision whether the given article, film, etc. is of interest or not. Keywords, categories, classes, etc. are also used in e.g., library classification systems and biological classification for helping the search and browsing amongst a large number of objects. In the latter cases the involved entities are categorised hierarchically, with a set of narrower or broader categories building up a tree-like structure composed of ``is a subcategory of'' type relations. In contrast, the tags appearing in on-line platforms are usually free words chosen by the author or owner of the given object, and tags are almost never organized into a pre-defined hierarchy of categories and sub-categories \cite{Mika_folk_and_ont,Spyns_folk_and_ont,Voss_cond_mat,our_ontology}. Furthermore, in many tagging systems like Flickr, CiteUlike or Delicious the tagging process is collaborative, as in principle an unlimited number of users can tag photos, Web pages, etc., with free words \cite{Cattuto_PNAS,Lambiotte_ct,Cattuto_PNAS2}. In order to highlight this collaborative nature, the arising set of free tags and associated objects are often referred to as folksonomies. Since the tagging actions involve user-tag-object triplets, a natural representations of these systems is given by hypergraphs \cite{Lambiotte_ct,Newman_PRE,Caldarelli_PRE,Schoder_tags,Zhou_recommend_overview}, where the hyperedges connect more than two nodes together.

An interesting problem related to free tagging is to extract a hierarchy between the tags based on their co-occurrences on the tagged items \cite{Garcia-Molina,Lerman_constr_2,Schmitz_constr,Van_Damme_constr,Tibely_plosone}. The basic motivation is that the way users think about objects presumably has some built in hierarchy, e.g., ``pigeon'' is usually considered as a special case of ``bird''. Revealing the hidden hierarchy between tags in a folksonomy or in a tagging system in general can significantly help broadening or narrowing the scope of search in the system, give recommendation about yet unvisited objects to the user, or help the categorisation of newly appearing objects \cite{Zhou_recommend_overview,Kazienko_chapter}. Here we apply an improved version of a recent tag hierarchy extraction method \cite{Tibely_plosone} to keywords associated to on-line articles, collected from the portals of Spiegel Online, The Guardian, The New York Times and The Australian. The obtained hierarchies show very interesting differences, indicating that the methods for choosing keywords are based on rather different principles in the studied journals. 

The structure of the paper is as follows: in \nameref{s:meth}, the principles of hierarchy construction and comparison are shortly described. In \nameref{s:data} the empirical datasets are presented, in \nameref{s:res} we compare the obtained hierarchies with each other, and in \nameref{s:dis} results are discussed.

\section{Methods} \label{s:meth}

\subsection{Hierarchy construction}

We employ an upgraded version of a recent method \cite{Tibely_plosone}, which is based on two assumptions: 
\begin{itemize}
 \item tags positioned high in the hierarchy also have high centrality values in the tag-tag coappearance graph, 
 \item parent-child pairs coappear more frequently than expected from pure chance. 
\end{itemize}
According to the first assumption, the algorithm orders the tags by their centrality, then, for each tag (which become child) the parent candidates are collected. All tags with higher centrality are parent candidates of the child tag. Candidate parents are assigned a score, indicating the probability of the observed number of co-occurrences according to a random null-model. Using the second assumption, the final parent is the candidate with the highest score sum, where the sum runs over all descendants of the child tag. Note, that the algorithm builds up the hierarchy bottom up, starting from the leaves with lowest centrality. The full detailed description of the currently used version of the algorithm involving a couple of improvements is given in the Supplementary Information.

\subsection{Similarity of hierarchies} \label{s:sim}

Hierarchies are frequently represented by Directed Acyclic Graphs (DAGs), in which directed cycles are forbidden. However, children are allowed to have more than one parent in general. For simplicity, we have restricted the number of parents to one in the present analysis. A natural idea for comparing two DAGs is to compare the hierarchical relations, i.e., the sets of ancestor-descendant relationships \cite{Tibely_plosone, Fattore, Brandenburg, Palla_palgrave, partialorder}. Here we adopt the approach proposed in Ref.\ \cite{Tibely_plosone}, defining a similarity measure based on mutual information. We note that mutual information plays a central role also in the comparison method introduced in \cite{Caldarelli} for the related, but separate problem of comparing hierarchical community structures, (where only the lowest-level nodes in DAG actually exist in the input data-set). The DAG similarity measure we use can be formulated as follows \cite{Tibely_plosone}
\begin{equation} \label{eq:infsym}
 I_{\alpha, \beta} = \frac{ 2 \sum_{x=1}^{N_{\alpha \beta}} |d_{\alpha}(x) \cap d_{\beta}(x)| \cdot \ln \left( \frac{|d_{\alpha}(x) \cap d_{\beta}(x)|(N-1)}{|d_{\alpha}(x)|\cdot|d_{\beta}(x)|} \right) }{ \sum_{x=1}^{N_{\alpha}} |d_{\alpha}(x)| \ln \left( \frac{|d_{\alpha}(x)}{N-1} \right) + \sum_{x=1}^{N_{\beta}} |d_{\beta}(x)| \ln \left( \frac{|d_{\beta}(x)}{N-1} \right) }
\end{equation}
where $\alpha$ and $\beta$ are two DAGs, having $N_{\alpha}$ and $N_{\beta}$ tags from which $N_{\alpha \beta}$ are common, and $d_{\alpha}(x)$ is the set of descendants of $x$ in DAG $\alpha$. Equation \ref{eq:infsym} is 0 for independent DAGs and 1 for identical ones.

A further very closely related similarity measure that turned out to be useful in previous studies is given by the linearised mutual information (LMI) \cite{Tibely_plosone}, based on the fraction of links that have to be rewired in a randomisation procedure on $\alpha$ leading to a hierarchy $\alpha_{\rm rand}$ with the same NMI when compared to $\alpha$ as the $I_{\alpha,\beta}$. The formal definition of this measure is given as follows. Let $I(f)$ denote the average NMI obtained for a fraction of $f$ randomly rewired links, $I(f) = <I_{\text{original,rand}}>_f$. By projecting the NMI of the empirical case, $I_{e}$, to the $f$ axis using this function as
\begin{equation}
 f^* = I^{-1}(I_{e}),
\end{equation}
we receive the fraction of randomly chosen links to be rewired in the empirical case for obtaining a randomized hierarchy with the same NMI. Based on that we define the linearised
mutual information, (LMI) as
\begin{equation} \label{e:lin}
 I_{\text{lin}} = 1 - f^* = 1 - I^{-1}(I_{e})
\end{equation}
This quality measure corresponds to the fraction of unchanged links in a random link rewiring process, resulting in a hierarchy with the same NMI as the empirical value. (The reason for calling it ``linearised'' is that equation \ref{e:lin} is actually projecting $I_{e}$ to the linear $1-f$ curve).

\section{Data} \label{s:data}

We analyse four tagged datasets, obtained from online news portals. They contain tagged news items, covering a more than 2 years long time window, in the same period. The four sources are: Spiegel Online, The Guardian, The New York Times and The Australian.

\subsection{General observations}

There are a few observations which hold for all four datasets. For example, very long tags exist, more like headlines ("\texttt{Muntazer al-Zaidi: the Iraqi shoe thrower}"). Some of the tags form ``frozen'' cliques in the coappearance network, where each member of such a clique appear only together with the other members of the clique, e.g., \texttt{Haiti} and \texttt{Haiti Earthquake Disaster 2010}, \texttt{Diana} and \texttt{Princess of Wales}. Since members of a large clique have large centrality values, such tags will be placed to unwanted high positions by the first step of the hierarchy construction algorithm. Therefore we have considered such ``frozen'' cliques as single tags, which fits better to the assumed usage of tags. 

Some concepts are represented by two or more tags, where the same idea is expressed with different, but synonymous words, e.g., \texttt{Art} and \texttt{Arts}. These were left as observed, unless explicitly stated otherwise. Another problem is posed by the occurrence of very rare tags, that are usually names. 

In order to avoid misleading results due to the above observed problems, we have prefiltered the tags by requiring that each tag pair in the coappearance network has to occur on at least $r$ news items. The $r=1$ case corresponds to skipping the prefiltering. We set $r$ to its optimal value for each dataset by keeping the number of tags as high as possible and minimizing the number of misleading tags described above. Finally we note, that temporarily important topics can produce unexpected  co-occurrences (e.g., "\texttt{Japan}" -$>$ "\texttt{Fukushima Nuclear Catastrophe}" -$>$ "\texttt{Nuclear Power}").

\subsection{Spiegel Online}

The dataset is from April 2011 to January 2013. It contains 4802 news items and 388 tags. For the pre-filtering, minimum 1 common news item for each tag pair (i.e., no filtering) seems to be a good tradeoff between noise reduction and info loss. The dataset looks very well organised (e.g., there are only ~400 tags, general tags are used consistently, and there are only a few duplicated tags, long tags or ``frozen'' cliques).

\subsection{The Guardian}

The dataset is from November 2009 to January 2013, containing 55835 news items and 6797 tags. Pre-filtering needs minimum 3 news items (removes 2530 tags and 61 news items). Here we found several ad hoc tags (mostly names), that were used only once or a handful of times. We found synonymous tags, e.g., "\texttt{Middle East and North Africa}" and "\texttt{Middle East}", that will appear as two local roots of two branches in the DAG. These branches correspond to the same topic, thus divide the related tags between them.

\subsection{The New York Times}

The dataset reaches from November 2010 to January 2013. It contains 35736 news items and 23009 tags. Cliques are a huge problem here. There are 2902 ones, collapsing them removes about 6000 tags. Several cliques appear on numerous objects, therefore the minimum news item-filtering does not solve the problem automatically. Cliques also reach very large sizes: there is a 809-tag clique (may contain much more characters than a news item itself); after the minimum news items filtering, the largest one still consists of 44 tags -- as follows from the definition of cliques, these tags appear strictly together on each object. For the pre-filtering, minimum 5 news items were required, leaving finally 2981 tags (out of 23009). News items were much less affected, 31184 out of 35736 remained.

\subsection{The Australian}

Data is from December 2009 to January 2013. It contains 31501 news items and 79054 tags -- thus, there are much more tags than news items. Cliques are present, but have only 1-2 objects, so it is not a serious problem, the pre-filtering can solve it. Multiple synonyms occur on the same object very often -- e.g., "\texttt{Economist\_Paul\_Samuelson}" "\texttt{Paul\_A.\_Samuelson}" "\texttt{Paul\_Samuelson}". Another example is the set of synonyms for Barack Obama, which are: \texttt{Barack\_Obama}, \texttt{BARACK\_Obama}, \texttt{Obama}, \texttt{PRESIDENT\_Barack\_Obama}, \texttt{President\_Barack\_Obama}, \texttt{President\_Obama}, \texttt{US\_PRESIDENT\_Barack\_Obama}, \texttt{US\_President\_Barack}, \texttt{US\_President\_Barack\_Obama}, \texttt{barack\_obama}. Pre-filtering with minimum 5 news items leaves 1673 tags out of 79504. The news items are reduced from 31501 to 10550. The tags have relatively few objects, and not only due to the large number of very infrequent tags, e.g., even the prime minister has only 900 objects. Although there are very general tags like \texttt{community}, \texttt{committee} or \texttt{claim}, most of tags are very specific, almost tailored for one object, e.g., \texttt{rebels\_storm\_Gaddafi\_compound}.

\section{Results}	\label{s:res}

We analysed the tag hierarchies obtained from an improved version of ``algorithm B'' published in Ref.\ \cite{Tibely_plosone}; a brief description of the idea of the method can be found in \nameref{s:meth}, the full details of the used algorithm are given in the Supplementary Information. In \nameref{s:indiv} first we summarise the most important properties of the individual hierarchies corresponding to the different news portals, which is followed by the pairwise comparisons in \nameref{s:pairw}. Finally, in \nameref{s:quant} we examine the overall quality of the hierarchies from different aspects.

\subsection{Analysis of the individual tag hierarchies} \label{s:indiv}

\begin{figure}[!h]
\begin{center}
\includegraphics*[width=\textwidth]{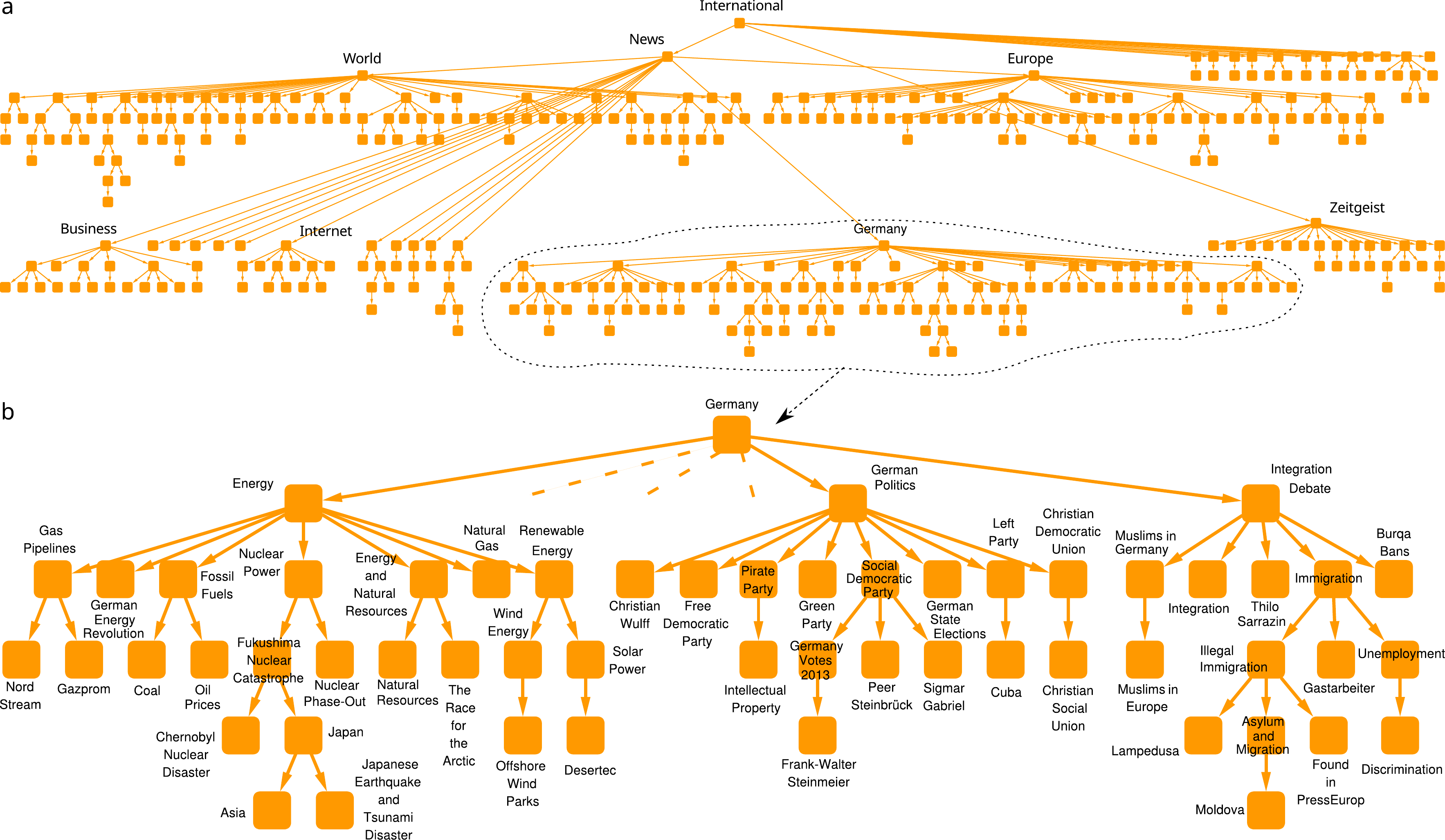}
\end{center}
\caption{Overview of the Spiegel DAG (top), and one part enlarged (bottom). The DAG is broken into two lines in the top figure to fit the whole graph in the available width. On the bottom figure, dashed lines indicate descendants which are not shown.}
\label{f:S_overw}
\end{figure}

\textbf{Spiegel Online} The constructed DAG consists of 1 connected component. Most of the tags are under 3 branches: "\texttt{World}", "\texttt{Europe}", "\texttt{Germany}". A visualisation of the DAG is shown on Fig.\ \ref{f:S_overw}.
The Spiegel DAG seems to be somewhat concerned with immigrants and integration, they have a branch containing $3.9\%$ of the tags,  similarly to Australian's $4.4\%$, and in contrast to $0.1\%$ and $0.7\%$ of Guardian and NYT (note that the latest data come from January 2013, well before the beginning of the recent migrant crisis).

\textbf{The Guardian} The overall structure of the DAG is quite well organised, the top 2-3 levels are very impressive. The DAG consists of four similarly-sized connected components: "\texttt{UK news}", "\texttt{World news}", "\texttt{Culture}", "\texttt{Sport}", although the tags "\texttt{World news}" and "\texttt{UK news}" are in isolated components, they are not completely mutually exclusive, e.g., both of them appear on the news items of "Defence policy". Note that while the components' top tags correspond well to the menu items on the journal's website, they are placed totally automatically by the DAG construction algorithm. Visualisation is omitted due to the relatively large size of the DAG, however, a smaller sample is shown in the Supplementary Information.
\begin{figure}[!h]
\begin{center}
\includegraphics*[width=\textwidth]{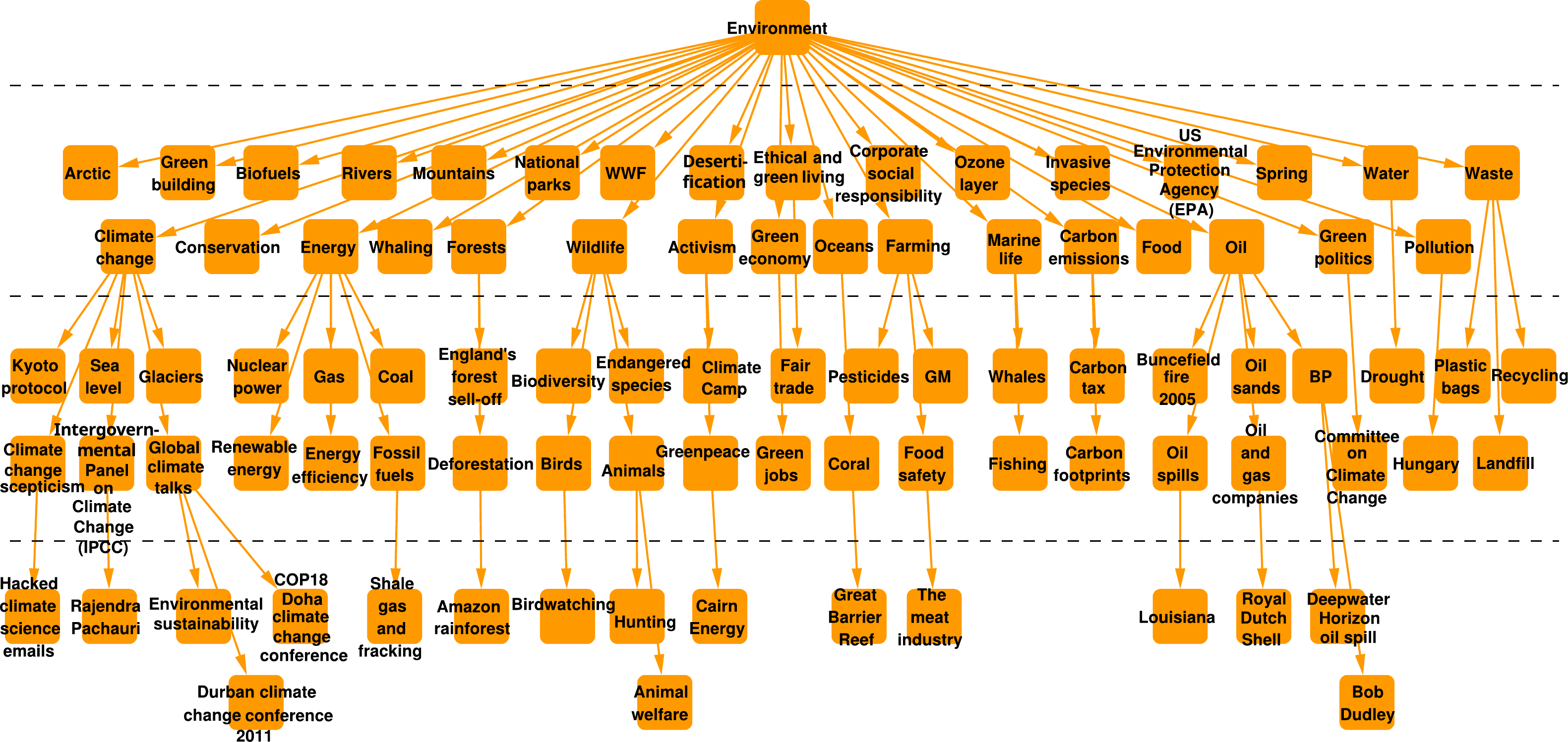}
\end{center}
\caption{Part of the Guardian's \texttt{Environment} branch, in the component \texttt{World news}. Hierarchical levels are separated by dashed lines.}
\label{f:G_env}
\end{figure}

\textbf{The New York Times} Here we found numerous duplicated branches in the constructed DAG (e.g., for research, television, education, medicine, defence and military forces). This indicates that for these topics, two distinct sets of tags were used in parallel. The DAG is much less organised than that of the Spiegel and of the Guardian. There are 31 isolated components, most of them correspond to one theme (e.g. "\texttt{Baseball}"). The sizes of the components varies from 898 to 2, and there is a continuous range of them from the 2nd largest one (274 tags) down. There are no very general categories. Although a number of large related components exists (under the tags "\texttt{Basketball}", "\texttt{Baseball}", "\texttt{Football}"), these components are not collected under a general "\texttt{Sport}" tag. It seems as if there were no demand for using general tags. Note that there is a tag called "\texttt{sports}", however, it appears only on 5 news items, and it is negligible. A technical consequence is that the DAG construction algorithm does not always select the most general tags as roots, because they lack the important connections to other components. Instead, one of the more specific tags can be selected for a central position, for example, "Middle East and North Africa Unrest (2010-)" for foreign affairs, or "\texttt{European Sovereign Debt Crisis (2010-)}" for Europe-related tags. In other words, the centrality no longer correlates only with the generality for the top tags.  Some lower-level branches end up at unexpected places, e.g., \texttt{Environment} under \texttt{Iran}. Superfluous levels appears, for example, \texttt{International Relations} under \texttt{United States International Relations}.

\textbf{The Australian} The DAG looks disorganised overall. There are about 1900 components for the 79504 tags without the pre-filtering, and about 300 components for the min.\ 5 news items-filtered 1673. There are no macroscopic components, the largest one's size is just 3480 (out of 79504 tags) and 165 (out of 1673 tags), which is less than 10\% of the total nodes. Even the existing components look more like just bunches of more or less associated tags than small hierarchical structures.

In general, the top of the constructed DAGs are much better than the bottom. This is no surprise - there is much more information for the construction algorithm at the top of the DAG.

\subsection{Pairwise comparisons} \label{s:pairw}

We carried out a pairwise comparison between the journals from the point of view of their content organisation. Since the audience and the interests of the journals are different, the list of tags appearing on the articles was unique for each news portal. Therefore, before actually comparing the tag hierarchies, first we needed to create a common tag set for each pair of journals. In a number of cases, finding the corresponding tag pairs went beyond a simple string matching and was based on semantic matching, e.g., "\texttt{Fossil fuels}" (Guardian) was matched with "\texttt{Oil (Petroleum) and Gasoline}" (NYT). The size of the reduced common tag sets were 252 (Spiegel-Guardian), 217 (Spiegel-NYT), 985 (Guardian-NYT), 93 (Australian-Spiegel), 278 (Australian-Guardian), 274 (Australian-NYT). 

The reduced hierarchies were obtained by keeping only the common tags in the original DAGs and erasing the rest of the tags. In most cases this resulted in deletion of leafs, sub-branches, or lower parts of sub-branches from the original hierarchies. However, a small number of times this procedure erased a tag higher in a given branch while keeping other tags lower in the same branch, therefore distorting the original DAG structure in a radical way. To ensure as much similarity to the original hierarchies as possible, under these circumstances the ancestors standing higher in the branch were also kept, despite that they were not part of the common tag set, (see the SI for more details). The reduced DAGs can be found in the SI.

\begin{figure}[!h]
\begin{center}
\includegraphics*[width=0.45\textwidth]{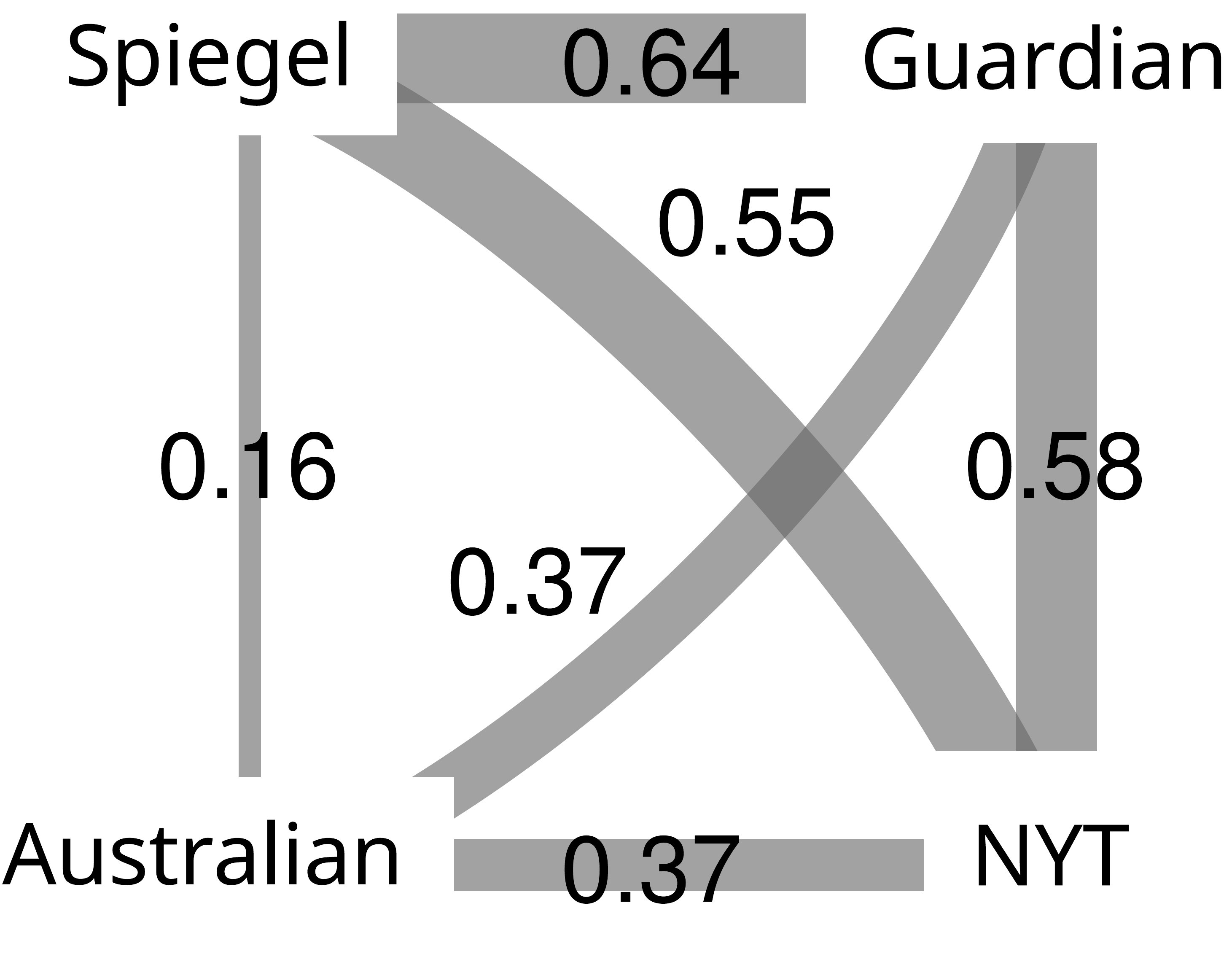}
\end{center}
\caption{Similarities between the news portals' DAGs, according to the mutual information-based linearised information similarity measure described in \nameref{s:sim}.}
\label{f:table}
\end{figure}

\begin{figure}[!ht]
\begin{center}
\includegraphics*[width=\textwidth]{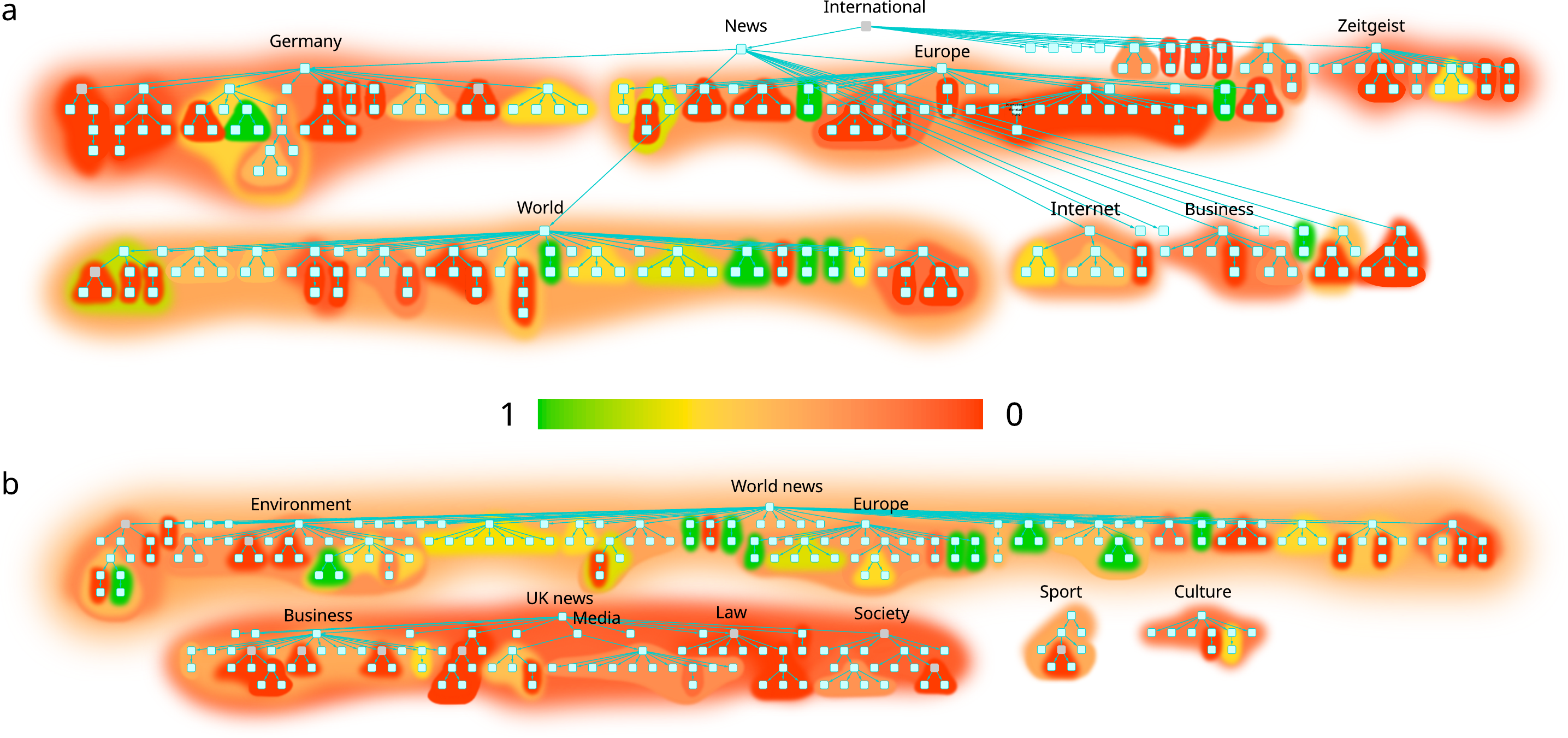}
\end{center}
\caption{The Spiegel (top) and the Guardian's (bottom) reduced DAG structures, providing the largest overall similarity in our analysis. For clarity, Spiegel's DAG is broken into two lines. Background colours show the result obtained by applying the similarity measure given in equation \ref{eq:infsym} to the given branch and the most similar branch from the other hierarchy. Note that sub-branches on all hierarchical levels have their own colour.}
\label{f:SG}
\end{figure}

For each pair of journals we have computed the linearised information similarity measure described in \nameref{s:sim} between the reduced DAGs, the obtained values are shown in Fig.\ \ref{f:table}. According to the results Spiegel and Guardian provide the largest similarity measure, which is also supported by a number of identical or almost identical sub-branches between the two DAGs, as shown in Fig.\ \ref{f:SG}. Here the background colouring of the sub-branches indicate the similarity to the corresponding (most similar) sub-branch in the other DAG.

\begin{figure}[!ht]
\begin{center}
\includegraphics*[width=\textwidth]{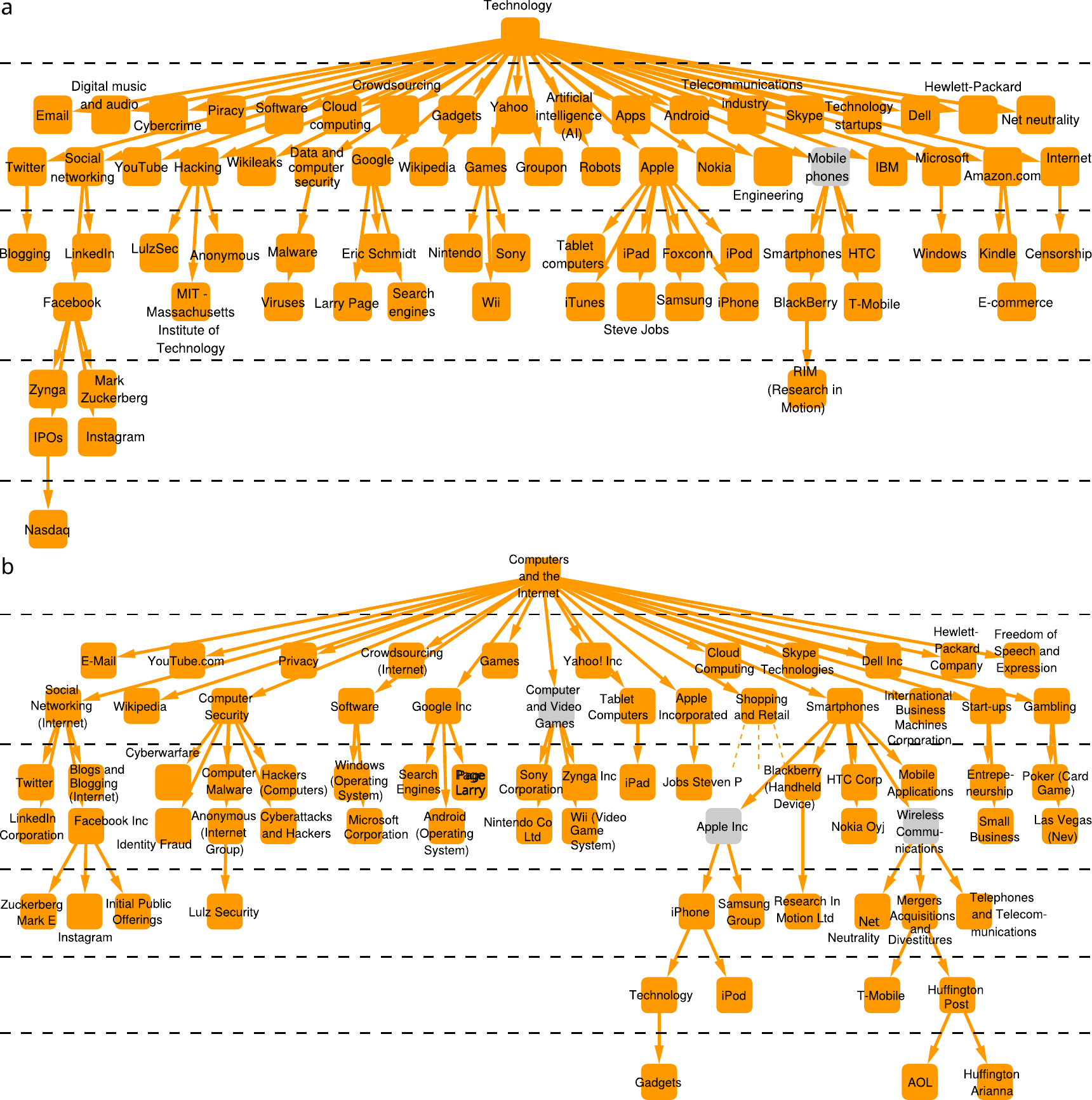}
\end{center}
\caption{Guardian's \texttt{Technology} branch (top) and New York Times' \texttt{Computers and the Internet} component (bottom). Hierarchical levels are separated by dashed lines. Grey tags do not appear in both DAGs, however, they connect branches containing common tags.}
\label{f:GN}
\end{figure}

The Spiegel, the Guardian and the New York Times have an overall similar structure, as Fig.\ \ref{f:table} shows, opposed to the Australian, which is dissimilar to all of them. Still, there are some differences between the first three journals. The Guardian, compared to the Spiegel, has a level of intermediately-sized branches, e.g., \texttt{law} or \texttt{society} in \texttt{UK news}. This level is missing from the DAG of Spiegel. Their global DAG structures are shown in Fig.\ \ref{f:SG}.
Meanwhile, the New York Times has interestingly no \texttt{World} tag, and foreign countries are separated into 4 different branches, in 3 components (see the SI for more details). Although the linearised information similarity between the Guardian and the New York Times is somewhat lower, they also have a few quite similar branches; a prominent example is shown in Fig.\ \ref{f:GN}.

\subsection{Statistical properties of the overall hierarchy structures} \label{s:quant}

According to the results presented in the previous sections the tag hierarchies obtained for the studied journals show strong differences. Here we examine to what extent does their overall structure follow a few simple intuitive requirements that can be formulated for a well organised tag hierarchy.

\textbf{Correlations with Google News.} One of the basic properties of a well organised hierarchy is that frequent, more general tags are expected to be higher compared to rare, specific tags. In order to examine the obtained hierarchies from this perspective we compared the centrality score of the tags in the tag co-occurrence network (determining their position in the hierarchy) with their number of hits provided by Google News. For each pair of tags with a significant number of co-occurrence we checked whether the difference between their centrality score and the difference between their number of hits in Google News have the same or the opposite sign. If the signs of the differences match for the majority of the tag pairs, then we can assume that the structure of the hierarchy is consistent with word frequencies of English news texts around the world.

In Table \ref{t:inversions} we show the relative frequency of the cases, where the differences have the opposite sign, calculated for tag pairs co-appearing in statistically significant numbers. If tags are assigned to articles absolutely at random, the result would correspond to a $0.5$ inversion rate, i.e., half of the coappearing tag pairs would have similar centrality and frequency ordering.
\begin{table}
 \begin{center}
 \caption{Ratios of inversions between centralities and real-world occurrence frequencies, calculated for tag pairs coappearing in statistically significant numbers. Totally random case corresponds to $0.5$.}
 \label{t:inversions}
 \begin{tabular}[ht!]{c | c}
  dataset & ratio of inversions\\
  \hline
  Spiegel  & $0.19$ \\
  Guardian & $0.21$ \\
  New York Times & $0.31$ \\
  Australian & $0.44$ \\
 \end{tabular}
 \end{center}
\end{table}
According to Table \ref{t:inversions}, the Spiegel and the Guardian data sets provide the best correspondence between tag frequency and centrality, with only a few percent difference in their score. They are followed by the New York Times, and finally, the Australian has a score close to the random case. Although the Google News data may be somewhat different from a fictitious collection word usage of all English speaking journalist, the results in Table \ref{t:inversions} show a quite clear-cut picture, which also corresponds well to the results of other comparisons.

\textbf{Geometrical properties of the hierarchies.} In this section we focus on the geometrical properties of the tag hierarchies from the perspective of whether their structure is helping navigability and search. First we examine the fragmentation of the DAGs, which we can quantify by first introducing the average size of the component of a randomly chosen tag given by,
\begin{equation} \label{e:s}
 \tilde{s} = \frac{\sum_i^{\text{tags}}s_i}{N}
\end{equation}
where $s_i$ is the size of the component containing tag $i$ and $N$ is the total number of tags. Based on $\tilde{s}$ we can calculate the expected lowest hierarchy level $l$ on which the top node of a branch of size $\tilde{s}$ would appear in a balanced $k$-ary tree of size $N$. In such a tree any branch can contain at most half of the tags of its mother branch, thus we define $l$ as
\begin{equation} \label{e:l}
 \begin{split}
 l &= \left \lceil \log_2 N / \tilde{s} \right \rceil, \quad \tilde{s} < N\\
 l &= 1, \quad\quad\quad\quad\quad\; \tilde{s} = N
 \end{split}
\end{equation}
where $\lceil x \rceil$ denotes the ceiling function of $x$. The value of $l$ becomes high for strongly fragmented tag hierarchies consisting of many small isolated components, where the navigability of the hierarchy is low. The results for $\tilde{s}$ and $l$ are summarised in Table \ref{t:l}. The tag hierarchy obtained for Spiegel (consisting of a single component) provides the lowest $l$ value, followed by Guardian and New York times. Apparently, the DAG of Australian is showing a very fragmented structure with $l = 6$.
\begin{table}
 \begin{center}
 \caption{Characteristic level showing the highest level of an idealised hierarchy to which an average connected component corresponds.}
 \label{t:l}
 \begin{tabular}[ht!]{c | r r c}
  dataset & $\tilde{s}\phantom{33}$ & $N$\phantom{3} & $l$\\
  \hline
  Spiegel  & 388\phantom{$.0$} & 388 & 1 \\
  Guardian & $1338.7$ & 4263 & 2 \\
  New York Times & $384.2$ & 2945 & 3 \\
  Australian & $46.2$ & 1487 & 6 \\
 \end{tabular}
 \end{center}
\end{table}

Another important question is whether branch sizes are balanced or not in the hierarchies. A well-balanced hierarchy is expected to have at least 2 but not more than $\mathcal{O}(1)$ comparably sized branches at every nonleaf tag. We define a balancedness measure with a pair of real numbers from  $[0,1)\times [0, 1)$ corresponding to the ratio of ``giant branches'' and the ratio of ``dwarf branches'' in order to quantify how a DAG fits to the above criterion. First, we calculate the cumulated size of the branches having a child branch which contains more than 50\% of the parent branch's tags. Second, we calculate the cumulated size of the child branches which are smaller than 10\% of their parent branches. The higher threshold is motivated by the fact that a child branch above 50\% is larger than all the other child branches combined. The motivation for the lower threshold is that below 10\%, for equal-sized child branches, the number of child branches exceeds $\mathcal{O}(1)$. Other numerical threshold values might also be applied, however, for demonstrating significant phenomena the precise value of the thresholds should not be important. We normalise the sums by their maximal possible value, thus, our balancedness measure is given by
\begin{equation}
 (R_g, R_d) = \left( \frac{\sum_g S_g}{\sum_b S_b}, \frac{\sum_d S_d}{\sum_b S_b} \right)
\end{equation}
where $b$ goes over all branches containing at least 2 tags, $S_b$ is the size of branch $b$, $g$ goes over branches containing a sub-branch having more than 50\% of $g$'s tags, and $S_g$ is the corresponding branch size, and $d$ goes over sub-branches which are smaller than 10\% of their parent branches with $S_d$ being the corresponding branch size.  A perfectly balanced hierarchy would have a (0,0) score and the two extremely unbalanced cases would have (1,0) for a chain and (0, 1) for a star graph. The results for $(R_g,R_d)$ are given in Table \ref{t:balance}.
\begin{table}
 \begin{center}
 \caption{Ratios of giant and dwarf branches among all branches, size-weighted.}
 \label{t:balance}
 \begin{tabular}[ht!]{c | c c}
  dataset & $R_g$ & $R_d$ \\
  \hline
  Spiegel  & $0.32$ & $0.22$ \\
  Guardian & $0.10$ & $0.42$ \\
  New York Times & $0.42$ & $0.22$ \\
  Australian & $0.26$ & $0.17$ \\
 \end{tabular}
 \end{center}
\end{table}

Spiegel's $R_g$ is dominated by a single contribution. The global root, \texttt{International} has a branch containing almost the whole DAG under \texttt{News}. Most of $R_d$ comes from small branches, although there are a few exceptions. In the Guardian DAG, dwarf sub-branches are common, due to the huge size of the components which dwarf several branches, as well as to nearly star-shaped branches, sometimes containing hundreds of leaf-tags (e.g., \texttt{Film}, \texttt{Music}). For the NYT, contrary to the Guardian, $R_g$ is much larger than $R_d$. Two important reasons are misplacing a number of branches and letting less general tags getting high centralities. Since the Australian DAG has quite limited structure inside the numerous small components, $R_g$ and $R_d$ are not very informative measures here. However, the tiny components seem to be well balanced. 

Further analysis of the DAGs can be found in the SI.

\section{Discussion} \label{s:dis}
We studied the hierarchy of keywords associated to news articles in four different on-line news portals. The datasets contain various artefacts, such as long and complex keywords, ``frozen'' cliques of exclusively coappearing tags, synonyms or very rare and specific tags. Nonetheless, it was possible for the construction method to obtain very reasonable DAGs from the data. The identification of frozen cliques might also be applied by disambiguation techniques, to identify cliques of equivalent semantic meaning, used in the field of Natural Language Processing. The constructed DAGs suggest that the tags appearing in the different news portals are organised to different degrees. Our analysis revealed that Guardian has an extra intermediate level of organisation at certain locations. A further very interesting result is that the number of connected components in the DAGs conveys information about the extent of organisation in the data: the Spiegel and Guardian have $\mathcal{O}(1)$ components and are quite organised, the New York Times has a few dozen components and breaks the world into independent pieces, and the Australian has $\mathcal{O}(100)$ components which are barely informative at all. 

A similar picture was emerging from the comparison between the frequencies of tags in Google News and their centrality score in the tag-tag co-appearance graphs. The correlation was quite strong in case of the Spiegel and the Guardian, medium for the New York Times, and almost equivalent to the totally random case for the Australian. A more detailed characterisation of the DAGs can be obtained by quantifying the extents of too large and too small sub-branches. Although being a geometry-based analysis, it can also identify problems with tag functions, like a non-comprehensive set of intermediate-level branches in the Guardian, or misplaced branches in the New York Times.

In summary, the following picture is arising from the different analyses we carried out: the Spiegel and Guardian datasets are quite well-organised, the New York Times is significantly less but still has relevant hierarchical structure, and the Australian is close to being random, from a hierarchical point of view. The consistency of the results is encouraging, and suggests that the measures used are useful in the quantification and comparison of datasets from the aspect of hierarchical organisation.

\section{Acknowledgements}
Financial support of the Hungarian National Science Fund (OTKA K105447) is acknowledged.

\section{Author Contributions}
GT, PP and GP conceived and designed the experiments. DSR provided the datasets. GT analysed the data. GT and GP wrote the manuscript. All authors reviewed the manuscript.

\section{Additional Information}
\textbf{Competing interests}: The authors declare no competing financial interests.

%
%
%
%

\end{document}